\documentclass[a4paper,fleqn,usenatbib]{mnras}

\usepackage{newtxtext,newtxmath}

\usepackage[T1]{fontenc}
\usepackage{ae,aecompl}

\usepackage{graphicx}	
\usepackage{amsmath}	
\usepackage{amssymb}	
\usepackage{soul}
\usepackage{amsbsy}     
\usepackage{float}
\usepackage{subfig}

\newcommand{\pval}{$p$-value$~$}
\newcommand{\pvals}{$p$-values$~$}
\newcommand{\chisq}{$\chi^2_{\rm{\mathcal N}}~$}
\newcommand{\pggpdf}{$P(\rm{G}|\rm{G}_{\rm{PDF}})$}
\newcommand{\plnlnpdf}{$P(\rm{LN}|\rm{LN}_{\rm{PDF}})$}
\newcommand{\pglnpdf}{$P(\rm{G}|\rm{LN}_{\rm{PDF}})$}

\newcommand{\pg}{$P(\rm{G})~$}
\newcommand{\pgpdf}{$P(\rm{G}_{\rm{PDF}})~$}
\newcommand{\pln}{$P(\rm{LN})~$}

\newcommand{\posln}{$P(+|\rm{LN})~$}
\newcommand{\negln}{$P(-|\rm{LN})~$}
\newcommand{\posg}{$P(+|\rm{G})$}
\newcommand{\negg}{$P(-|\rm{G})$}

\title[Non Gaussian PDFs in Time Series]{Deviations from normal distributions in artificial and real time series: a false positive prescription}

\author[P. J. Morris et al.]{
Paul J. Morris,$^{1}$\thanks{E-mail: paul.morris@physics.ox.ac.uk (PJM)}
Nachiketa Chakraborty,$^{2,3}$
and Garret Cotter$^{1}$
\\
$^{1}$Oxford Astrophysics, Denys Wilkinson Building, Keble Road, Oxford, OX1 3RH, United Kingdom\\
$^{2}$Max-Planck-Institut f\"ur Kernphysik, Saupfercheckweg 1, 69117 Heidelberg, Germany\\
$^{3}$Data Assimilation Research Centre, University of Reading, Whiteknights Rd, Reading, RG6 6BB, United Kingdom
}

\date{Accepted XXX. Received YYY; in original form ZZZ}

\pubyear{2019}

\begin{document}
\label{firstpage}
\pagerange{\pageref{firstpage}--\pageref{lastpage}}
\maketitle





\begin{abstract} 
Time series analysis allows for the determination of the Power Spectral Density (PSD) and Probability Density Function (PDF) for astrophysical sources. The former of these illustrates the distribution of power at various timescales, typically taking a power-law form, while the latter characterises the distribution of the underlying stochastic physical processes, with Gaussian and lognormal functional forms both physically motivated. In this paper, we use artificial time series generated using the prescription of Timmer \& Koenig to investigate connections between the PDF and PSD. PDFs calculated for these artificial light curves are less likely to be well described by a Gaussian functional form for steep ($\Gamma \gtrapprox 1$) PSD indices due to weak non-stationarity. Using the {\sl Fermi} LAT monthly light curve of the blazar PKS2155-304 as an example, we prescribe and calculate a false positive rate which indicates how likely the PDF is to be attributed an incorrect functional form. Here, we generate large numbers of artificial light curves with intrinsically normally distributed PDFs and with statistical properties consistent with observations. These are used to evaluate the probabilities that either Gaussian or lognormal functional forms better describe the PDF. We use this prescription to show that PKS2155-304 requires a high prior probability of having a normally distributed PDF, \pg $\geq~ 0.82$, for the calculated PDF to prefer a Gaussian functional form over a lognormal. We present possible choices of prior and evaluate the probability that PKS2155-304 has a lognormally distributed PDF for each. 

\end{abstract}




\section{Introduction}\label{sec:intro}


Time series analysis is crucial to study periodic and transient phenomena in astrophysics. Rapid X-ray variability is a common feature in the time series of compact accreting objects such as Black Hole X-ray Binaries (BHXRBs) \citep{1973A&A....24..337S} and Active Galactic Nuclei (AGN) \citep{1995PASP..107..803U}, which also display rapid variability in high energy $\gamma$-rays \citep[e.g.][]{2013ApJ...773..177N}. The physical origin of this variability is likely related to underlying physical processes occurring in the object which can leave an imprint in the time series properties \citep{2018MNRAS.480L.116S}. 

One common property of astrophysical time series used to examine variability in astrophysical time series is the Power Spectral Density (PSD), which quantifies the amount of power in given frequencies sampled by the time series. This is typically estimated by utilising the method of \citet{1948Natur.161..686B}, where the time series is divided into $M$ non-overlapping sections for which the discrete Fourier transform (DFT) is calculated, and the PSD is taken as the mean of the DFTs calculated for each individual section. The functional form of the PSD is often approximately that of a power law, which may include a characteristic break separating regions of different spectral indices \citep[e.g.][]{2002MNRAS.332..231U}. 


Another property of a time series is the Probability Density Function (PDF), which can be obtained by forming a histogram of the flux values of an individual light curve. A PDF effectively quantifies the probability of a particular source being observed at a given flux value. If a time series is represented by a model containing an additive sequence of independent random variables, then it is said to be linear. Here, the time series $y$-values are expected to be normally distributed via the central limit theorem, where $y$ represents a desired measured quantity such as flux. Accordingly, the PDF is well described by a Gaussian functional form. If the functional form of the PDF deviates from a Gaussian this may be indicative of underlying physical processes occurring in the source of interest. One such distribution known to occur in nature is the lognormal distribution, which can occur in the case of the random constituent elements in a time series elements being multiplicative rather than additive \citep[e.g.][]{UMV:2005,Heyde:2010}. 

PDFs have increasingly been computed for astrophysical sources \citep{UMV:2005}, with lognormality ubiquitous in BHXRBs \citep{2004A&A...414.1091G,2012MNRAS.422.2620H} and also inferred from PDFs calculated by flux binning several blazar light curves first in X-rays \mbox{\citep{2009A&A...503..797G}} and optical \citep{2018ApJ...857..141S} and later in GeV gamma-rays \citep{2017ApJ...849..138K} and across the electromagnetic spectrum \citep[e.g.][]{2155logN, 421logN}. Though these characteristics have been speculated as applying to all accretion powered compact objects exhibiting rapid variability \citep[e.g.][]{UMV:2005}, the physical origin, and whether such behaviour generally applies to the blazar population, is currently unknown. The physical motivation for lognormally distributed time series in BHXRBs arises from accretion disc flicker models \citep[e.g.][]{1997MNRAS.292..679L,2001MNRAS.327..799K,2004MNRAS.348..111K,2006MNRAS.367..801A}. Here, fluctuations of the accretion rate propagate inwards through the disc causing the modulation of faster fluctuations at smaller radii. Fluctuations in the X-ray flux are proportional to variations in the accretion rate, resulting in the source PDF being accurately described by a lognormal function. Observations of long-term lags in optical emission relative to X-rays are consistent with this model as optical and X-ray emitting regions of the accretion disc are modulated together \citep{2009MNRAS.394..427B}. It has been further speculated that lognormal PDFs from AGN jets could be indicative of a disc-jet connection \citep{2008bves.confE..14M}. 


In this paper, we use simulations of artificial time series to investigate the fundamental relationship between the PSD and PDF, both of which are important properties of astrophysical sources. We begin by investigating the impact on a steepening PSD spectral index on the functional form of the PDF, before looking at how changing the minimum and maximum frequencies sampled in the light curve affect the PDF. We then use these results to motivate a prescription for a false positive fraction, which quantifies the probability of incorrectly measuring the wrong PDF from a source in which the intrinsic PDF has a different functional form. 
Specifically, we consider the case where we the true source PDF can only be either Gaussian or lognormal, and generate many artificial time series for each of these known PDF functional forms. We base our light curve time sampling on ${\sl Fermi}$ Large Area Telescope (LAT) monthly light curves for our artificial blazar time series and incorporate the measured PDF spectral index into our false positive calculation. We conclude by computing the false positive posterior probability for determining the correct PDF functional form for the blazar PKS2155-304 as an example. 

\section{Time Series Simulations}\label{sec:TimeSim}

Observed time series are necessarily discrete, which leads to difficulties computing the PSD. These include red noise leakage \citep{1978IEEEP..66...51H}, the effects of which can be mitigated by simulating a longer light curve ($\geq 10$ times the desired time series length) such that power is contained in frequencies lower than the minimum sampled by the time series \citep{2002MNRAS.332..231U}. The finite time resolution can also lead to aliasing effects \citep[e.g.][]{PhysRevE.71.066110}. This occurs when frequencies higher than the sampling frequency are present in the time series and add additional power to lower frequency elements and is prevented by Nyquist sampling where the data is sampled at at least twice the desired maximum. When producing artificial time series, the effects of aliasing can be avoided by generating time series from a known PSD, in which the resultant light curves are sampled at the desired Nyquist frequency \citep{2002MNRAS.332..231U}. Artificial, user-defined, time series alleviate further issues that may affect data time series such as uneven sampling or variable signal to noise levels. Additionally, time-series simulations can be used to generate an ensemble of ``realistic" lightcurves i.e. those matching the cadence of observations and with comparable statistical moments. This ensemble can be used to provide statistical tests including computing the significance of a test; an example of this is found in \citet{2018Galax...6..135R}. Presently there are two main prescriptions used to simulate artificial light curves which are outlined in the following subsections. 


\subsection{PSD Functional Forms}\label{sec:PSDforms}

In the GeV $\gamma$-ray regime, the BL Lacs and FSRQ sub-classes of AGN, which are those which have their relativistic jets pointed along a sight-line close to that of the Earth, have PSDs well described by a single power law with average spectral indices of 1.7 and 1.5, respectively \citep{2010ApJ...722..520A} (see also \citet{2013ApJ...773..177N}). Accordingly, one method of generating artificial time series relies on the definition of a power law PSD, $\mathcal{P}_{\rm{PL}}(\nu)$, which quantifies the power per unit frequency at frequencies $\nu$. This is defined as,
\begin{equation}
  \mathcal{P}_{\rm{PL}}(\nu) = A\nu^{-\Gamma}
  \label{eq:PSD}
\end{equation}
for some normalisation $A$ and power law index $\Gamma$. Alternatively, many compact object sources have been shown to have a broken power law (BPL) PSD, which has been calculated using X-ray observations of BHXRBs \citep[e.g.][]{0004-637X-514-2-939,2003A&A...407.1039P,2005A&A...440..207B} and AGN \citep[e.g.][]{Marshall:2015}, which can also have broken power law PSDs in the optical \citep{2018ApJ...857..141S}. These PSD typically begin with a white noise index of $\Gamma=0$ which transitions to pink or red noise at a characteristic break frequency which may be related to a characteristic cooling timescale of the emitting population \citep[e.g.][]{2012A&A...540L...2I}. 

The functional form of a BPL PSD is defined as,
\begin{equation}
  \mathcal{P}_{\rm{BPL}}(\nu) = 
    \begin{cases}
     A \nu^{-\Gamma_1} & \nu \leq \nu_{\rm{b}}\\
     A \nu_{\rm{b}}^{(\Gamma_2-\Gamma_1)}\nu^{-\Gamma_2} & \nu > \nu_{\rm{b}}\\
    \end{cases}
    \label{eq:BPL}
\end{equation}
transitioning in spectral index at a given break frequency, $\nu_{\rm{b}}$. For observed time series, the PSD index often flattens towards the low frequency end of the power spectrum due to finite observation length and sampling of the time series \citep{2012A&A...540L...2I}.

\subsection{Timmer-Koenig Simulations}\label{sec:TK95}

A popular prescription for the generation of simulated time series is the method of Timmer and K\"onig \citep[][hereafter TK95]{TK:95}. TK95 simulations essentially generate artificial lightcurves with power-law noise that are easily extendable to other user-defined forms and have Gaussian PDFs. The method starts assuming a power spectral shape describing a time-series with a general power-law type noise such as white (0.0), pink (1.0) or red (2.0) noise. Using this, real and imaginary parts of Fourier amplitudes are drawn from a Gaussian distribution with a normalisation such that the variance is that of the observed (or user defined) lightcurve. These Fourier amplitudes are used to re-construct the time-series which is now a realisation of the underlying distribution of the physical process driving variability that we wish to probe. As stated in TK95, this ensures that we have simulated lightcurves that preserve the observed variability properties to 1st order (as mean and variance is matched with observations). In doing so, as usual the length or duration of the simulated lightcurve is a factor $\geq$ 10 longer than the observed lightcurves to avoid red noise leakage or loss of power in the longer timescales.

The discrete inverse Fourier transform (DiFT) of this artificial PSD yields a simulated data set with a PSD consistent with the desired user-defined PSD. This method can also be used to produce artificial light curves with the same PSD as a given data set by performing a discrete Fourier transform (DFT) and randomising the amplitudes and phases in Fourier space before calculating the time series via an DiFT. The TK95 method is popular because of its simplicity, and the artificial time series it produces, $y_{\rm{TK95}}(t)$, have PDFs with fluxes that are distributed normally. This property also allows lognormally distributed time series to be produced by exponentiating the output normally distributed time series, i.e. $y_{\rm{LN}(t)} = \exp(y_{\rm{TK95}}(t))$ \citep{UMV:2005}. Unless otherwise explicitly stated, artificial time series used in this work have been generated using an input power-law PSD as defined in Eqn. \ref{eq:PSD} and using the method outlined below in Section \ref{sec:TK95}.

\begin{figure}{\includegraphics[width=\columnwidth]{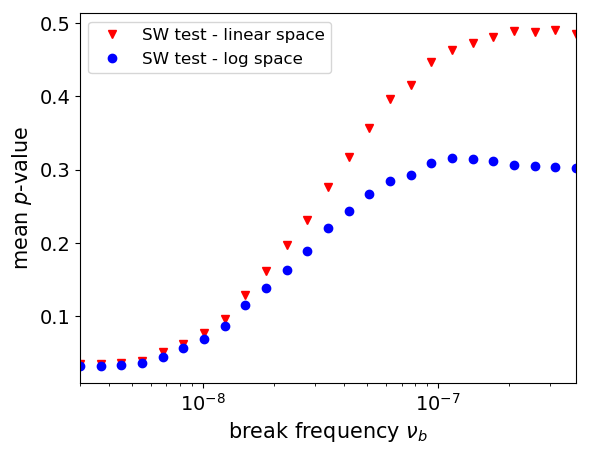}} \caption{Figure showing the evolution of the $p$-value returned from a Shapiro-Wilk test (see Section \ref{sec:tests}) for an ensemble of 10,000 TK95 simulated time series generated with a broken power law PSD transitioning from white to red noise spectral indices at a variable frequency, $\nu_{\rm{b}}$. The figure shows the dependence of the mean \pval on $\nu_{\rm{b}}$. Towards the left hand side of the plot, the PSD spectral break is at low frequencies, increasingly reducing the PSD to a pure red noise power law. Towards the right hand side, a white noise PSD is recovered as the simulated time series have fluxes consistent with being normally distributed. The red triangles indicate tests for normality in linear space, whilst the blue circles represent tests for normality in logarithmic space.}
\label{fig:pvalVbreakfq}
\end{figure}

\subsection{Emmanoulopoulos Simulations}\label{sec:EMP13}

To generate a time series with a known PDF and PSD, the more sophisticated Emmanoulopoulos \citep[][hereafter EMP13]{2013MNRAS.433..907E} method can be used. This method combines features from TK95 and that of \citet{PhysRevLett.77.635}. It works by initially generating two time series of equal length; one from the TK95 method, $x_{\rm{TK95}}(t)$, and the second is drawn directly from the desired PDF, $x_{\rm{PDF,i}}(t)$. The discrete Fourier transforms of both of these time series are taken, with the amplitudes from the TK95 DFT combined with the phases from the PSD obtained from the DFT of $x_{\rm{PDF,i}}(t)$ to give a modified PSD. The DiFT of this gives an adjusted time series, $x_{\rm{adjusted,i}}(t)$. Elements in $x_{\rm{adjusted,i}}(t)$ are then replaced by those from $x_{\rm{PDF,i}}(t)$ according to the ranking of the former to give a new time series, with the desired PDF, $x_{\rm{PDF,i+1}}(t)$. If $x_{\rm{PDF,i+1}}(t) \neq x_{\rm{PDF,i}}(t)$, then $x_{\rm{PDF,i+1}}(t)$ replaces $x_{\rm{PDF,i}}(t)$ until convergence of the two is found. This results in an artificial time series with both the desired PDF and PSD. This method is more computationally intensive than the TK95 method because of the iterative element.

\begin{figure*}
\centering
    \subfloat[]{\includegraphics[width=1.0\textwidth]{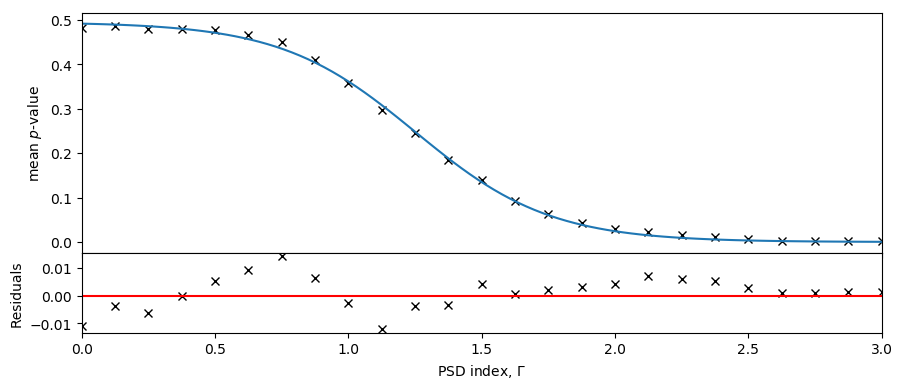}} \label{SEDfit} \\

\subfloat[]{\includegraphics[width=1.0\textwidth]{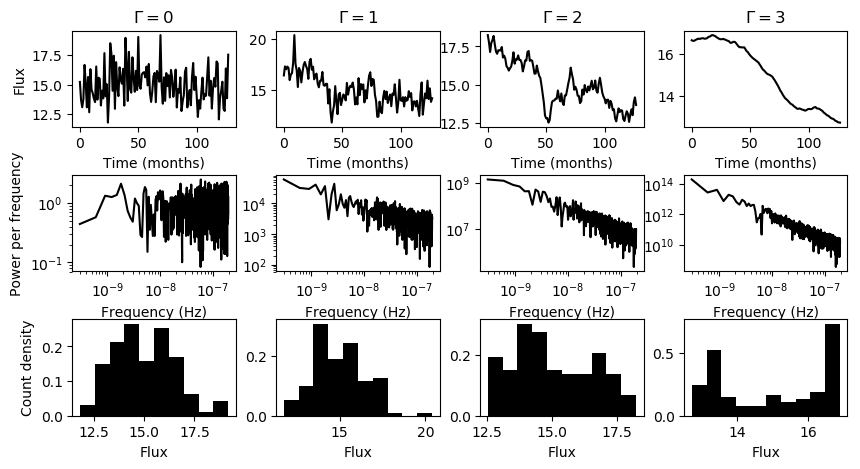}}\label{parflare} 
\caption{\small (a) Figure demonstrating the effect of varying the PSD index on the $p$-value for a Shapiro-Wilk test assuming a power law noise PSD. The artificial TK95 generated light curves in this calculation were 128 months long and binned in one month time intervals. It can be seen that the $p$-value sharply decreases at around $\Gamma = 1$, reducing the number of artificial time series which are consistent with having normally distributed time series. Above $\Gamma \approx 2$, the null hypothesis can be rejected to $> 2\sigma$. The mean $p$-values are calculated using 10,000 simulations for each PSD index. The fiducial fit representing the blue solid line in the top panel is described by Eqn. \ref{eq:func}. (b) Sample time series, PSDs and PDFs for a sample of power law spectral indices used in the top panel. It can be seen that the light curves become increasingly dominated by the low frequency components, which can lead to the PDF deviating from a Gaussian functional form. The PDFs here have been generated using 10 uniform bins in linear space. }
\label{fig:PSDev}
\end{figure*}

\begin{figure*}
{\includegraphics[width=\textwidth]{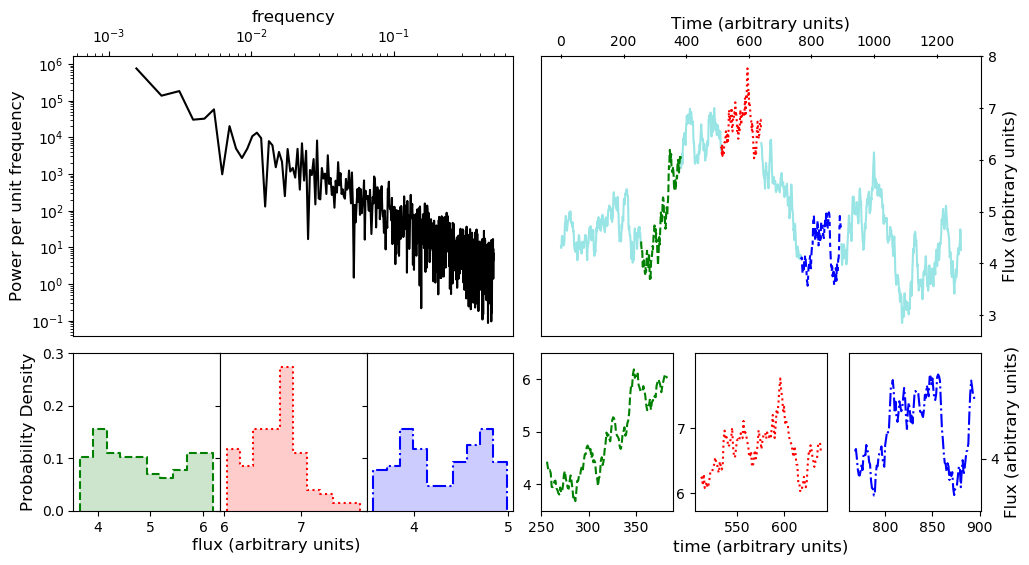}} 
\caption{Figure demonstrating the non-stationarity of time series for a power law PSD with index $\Gamma=2$. The top left panel shows the user-defined long-lightcurve PSD, while the top right shows the corresponding long artificial light curve generated from it. The coloured sections (green dashed, red dots and blue dot-dashed) in the long time series, which are representative of TK95 generated light curves, are plotted in the bottom right subplots. Their corresponding PDFs are shown by the colour and border style matching histograms below the PSD. It can be seen that light curves sampled from the same long time series can have PDFs that differ vastly in functional form, indicating a lack of stationarity. This highlights the difficulty associated with correctly measuring the PDF for an object whose light curve is only sampled over a relatively short amount of time.} 
\label{fig:nonStat}
\end{figure*}

\section{Tests for normality}\label{sec:tests}

We aimed to initially establish the likelihood of simulated TK95 time series being consistent with having normally distributed fluxes. A metric to asses the normality of a time series was therefore required, with three popular methods considered.

Commonly in astronomy literature, the PDFs are estimated from the lightcurves (time-series) as a histogram with fixed uniform binning to obtain a PDF in histogram form \citep[e.g.][]{2017ApJ...849..138K,2018RAA....18..141S}. The best fit to this PDF is then assessed used a variety of standard models such as Gaussian and lognormal distributions, with the goodness of fit evaluated with a standard statistical test such as the reduced chi squared statistic, $\chi_{\mathcal N}^2$ \citep[e.g.][]{HandH}, where $\mathcal N$ is the number of degrees of freedom. The model with the best $\chi_{\mathcal N}^2$ value, i.e. $\chi_{\mathcal N}^2 \approx 1$, is considered to best fit the histogram. This method is further discussed in Section \ref{sec:falsePos}. However, because the histogram has a dependency on the binning algorithm we decided to include additional statistical tests.  

A common test used to assess the likelihood of a time series being normally distributed in astronomy literature \citep[e.g.][]{2017ApJ...849..138K} is the Anderson-Darling (AD) test \citep{anderson1952}. The AD test can be used to determine how likely a sample is to come from a specified distribution, however, Monte Carlo simulations have indicated that when assessing if a sample is consistent with being normally distributed, the AD test has less statistical power than a Shapiro-Wilk test \citep{normtests,normtests2}. We therefore instead use the Shapiro-Wilk (SW) \citep{SWtest} as our test for time series normality. This test can also be used to test for lognormality by taking the natural logarithm of each time series and testing for normality in log space. Furthermore, the SW test is known to be reliably applicable up to relatively large data sample sizes of $\leq 5000$ \citep{Mahibbur:1997}, thus is suitable for assessing whether the comparatively shorter time series simulations here are normally distributed.

An important property of the SW test considered in this work is the $p$-value. This gives the probability of obtaining the data under the condition that the null hypothesis is true. The null hypothesis for an SW test is that the sample is  normally distributed. We use the $p$-value to illustrate the average consistency of our TK95 simulations to be able to produce normally distributed time series.

\section{\protect\boldmath Analytical Motivation for a Transition at $\Gamma=1$}\label{sec:analytic}

It has been noted that properties of TK95 simulated time series vary as a function of PSD index. \citet{2003MNRAS.345.1271V} show that the distribution of variances of these time series transition from a Gaussian for $\Gamma = 0$ and approaches a $\chi^2$ distribution as the index becomes steeper. The interpretation of this is that the steepening PSD spectral index progressively diminishes the effect of high frequency components, effectively reducing the total number of degrees of freedom in each simulated time series \citep[see also][]{UMV:2005}. More recent work by \citet{2019MNRAS.485..260A} has shown that the exponential flux distributions at steep PSD indices can deviate from the expected lognormal distribution.

Analytically, this can be explained by first obtaining the total power in a time series by integrating the PSD (Eqn. \ref{eq:PSD}) with respect to frequency,
          
\begin{equation}
        \int_{\nu_{\rm{min}}}^{\nu_{\rm{max}}} \mathcal{P}(\nu) ~{\rm d}\nu =  \int_{\nu_{\rm{min}}}^{\nu_{\rm{max}}} A\nu^{-\Gamma} ~{\rm d}\nu 
              = \frac{A\nu^{1-\Gamma}}{1-\Gamma} \biggr\rvert_{\nu_{\rm{min}}}^{\nu_{\rm{max}}}  
        \label{eq:intPSD}
\end{equation}
It can be seen from Eqn. \ref{eq:intPSD} that the pink noise case when $\Gamma=1$ is unique as the denominator causes both integration limits to diverge. This means that any finite time series, which by definition has discrete frequencies, contains less power at either end of the PSD than an ideal pink noise power spectrum. 

When $\Gamma<1$, $(1-\Gamma)>0$. From Eqn. \ref{eq:intPSD}, this causes the high frequency terms to be significant and approach $\infty$ as $\nu_{\rm{max}} \rightarrow \infty$. This means that the PSD for a finitely sampled time series will contain less power for $\Gamma<1$ than a pink noise PSD at higher energies. In contrast, the low frequency limit will reduce to zero as $\nu_{\rm{min}} \rightarrow 0$.  This is important because it shows that when $\Gamma<1$, the high frequency PSD components are able to provide a significant contribution to the total power in the time series.

Conversely, if $\Gamma>1$, $(1-\Gamma)<0$. Eqn. \ref{eq:intPSD} shows that as $\nu_{\rm{max}} \rightarrow \infty$, the contribution from high frequency terms approaches zero, so the relative power in them diminishes at steeper PSD indices and is finite for steep PSD indices. Here, the low frequency limit approaches $\infty$ as $\nu_{\rm{min}} \rightarrow 0$. It is therefore for $\Gamma \gtrapprox 1$ that the PSD and time series become dominated by increasingly low frequency components which contain a substantial fraction of the total power. If the PSD index is steep enough, the time series will be dominated by a small number of low frequency components, which we show later can yield time series which depart significantly from normally distributed PDFs.

Therefore, $\Gamma \approx 1$ is roughly where the contribution to the PSD from the highest sampled frequencies becomes sub-dominant relative to their low frequency counterparts. Their influence on the behaviour of the artificial light curves begins to diminish, progressively reducing rapid scale variability for high PSD indices.  Recent work by \citet{2019MNRAS.485..260A} has shown that mis-measurement of the PSD can affect the functional form of the PDF. We undertake simulations that look into the connection between the PDF functional form and PSD spectral index and quantify the impact of this effect on any deviation of simulated light curves from normally distributed fluxes in the following sections.

\section{Method and Results}\label{sec:method}

Initially, we wished to demonstrate the analytic result presented in Section \ref{sec:analytic}, and test the relative contribution of the high frequency PSD components towards the properties of the PDFs for artificial time series generated with different PSD spectral indices. We decided to use a BPL PSD as defined in Eqn. \ref{eq:BPL} to restrict the power in the higher frequencies above the break frequency, $\nu_{\rm{b}}$. As we analytically expected any transition to occur around $\Gamma = 1$, we defined our BPL PSD to transition from white to red noise such that $\Gamma_1 = 0$ and $\Gamma_2 = 2$ in Eqn. \ref{eq:BPL}. We varied the value of $\nu_{\rm{b}}$ logarithmically uniformly throughout the entire frequency range of the PSD. For each value of $\nu_{\rm{b}}$, 10,000 light curves were simulated using the TK95 method outlined in Section \ref{sec:TK95}, which was chosen over the EMP method as the latter requires significantly more computation time as a consequence of the iterative element. To mirror {\sl Fermi} LAT blazar time series, these simulated data had their timing properties taken assuming monthly time bins over the $\approx 10$ year total time observations have been made using the LAT. Our artificial time series were therefore set up to have 128 elements with bin size equal to one month. In this test, we assessed the normality of simulated time series with a $p$-value returned from a Shapiro-Wilk test, and the lognormality by testing normality in log space with the same test. The null hypothesis is that data are normally distributed. 

Fig. \ref{fig:pvalVbreakfq} shows that the $p$-value is close to zero for $\nu_{\rm{b}} \approx \nu_{\rm{min}}$ when the PSD is approximately pure red noise and increase before converging to $\sim 0.5$ as $\nu_{\rm{b}} \rightarrow \nu_{\rm{max}}$ (approaching white noise), thus demonstrating a transition occurs between $\Gamma=0$ and $\Gamma=2$. Intermediate points shift $\nu_{\rm{b}}$ and the transition point between the two. Fig. \ref{fig:pvalVbreakfq} also shows that the highest frequencies can significantly contribute towards a time series with a white to red noise PSD if the break frequency is greater than the mean logarithmic frequency of the PSD. It also infers that the characteristic shape of the PDF is not consistent with being Gaussian in functional form for PSDs with a substantial red-noise spectral component. This is because a low $\nu_{\rm{b}}$ effectively makes the PSD close to pure red noise, lowering the relative contribution from the high frequency components and reducing the total number of degrees of freedom, as discussed in \citet{2003MNRAS.345.1271V}. The figure also indicates greater average consistency with the null hypothesis being true for Gaussian PDFs over lognormal for $\nu_{\rm{b}} \gtrapprox 10^{-7}~\rm{Hz}$, yet it should be noted that in this frequency range the null hypothesis cannot be rejected to a significant confidence level in either case as the $p$-value is above the 2$\sigma$ threshold of 0.05. The lower average \pval for the SW test for a normally distributed PDF in log-space relative to in linear-space indicates that the y-values of the artificial time series are inconsistent with being normally distributed in log-space for a greater proportion of realisations than those with PDFs that are inconsistent with being normally distributed in linear space.

Secondly, it was decided to investigate the effect that changing the PSD index has on the likelihood of obtaining normally distributed artificial time series. In this test, for the sake of simplicity, it was assumed that simple power law as in Eqn. \ref{eq:PSD} accurately quantifies the PSD, as appears to be the case for blazars \citep[e.g.][]{2010ApJ...722..520A,2013ApJ...773..177N}. The PSD index for blazars is typically in the range $\Gamma \approx 1-2$ \citep[e.g.][]{2010ApJ...722..520A,2013ApJ...773..177N,2014ApJ...786..143S}, so this was included in our range which spanned from white noise ($\Gamma = 0$) to $\Gamma = 3$. 

For each PSD index, 10,000 light curves were generated using the TK95 method outlined in Section \ref{sec:TK95}. We find that the results show high consistency for a sample size of $\gtrapprox 100$ artificial time series generated at each value of $\Gamma$. The statistical properties of each time series were evaluated in accordance to the statistical tests outlined in Section \ref{sec:tests}, with a $p$-value returned for the SW test. Our main findings may be summarised as follows:

\begin{itemize}
    \item As expected, a transition occurs for $\Gamma>1$, with the average $p$-value decreasing for steeper PSD indices, for TK95 simulated time series. For $\Gamma>1$, there is a sharp decline in the average $p$-value, thus these artificial time series are more likely to be inconsistent with the null hypothesis of having normally distributed fluxes. This is shown in Fig. \ref{fig:PSDev} with the mean $p$-value from SW tests plotted against $\Gamma$.
    \item The functional form of the curve described above is approximately that of a sigmoid, and is discussed in Section \ref{sec:func}. 
    \item At steeper PSD indices, especially above $\Gamma>2$, the time series PDF is, on average, not obviously consistent with a physically motivated PDF functional form such as a Gaussian or lognormal. 
\end{itemize}

Fig. \ref{fig:PSDev} shows the dependency on the mean \pval on PSD index. By comparison with Fig. \ref{fig:pvalVbreakfq}, it can be seen that the \pval level for white noise in Fig. \ref{fig:pvalVbreakfq} is equal to the \pval before $\Gamma=1$ in Fig. \ref{fig:PSDev}, thus in this region on average there is little difference to time series generated from white noise PSDs. Fig. \ref{fig:PSDev} also offers a qualitative explanation which explains why the PDF functional form becomes consistent with being non-Gaussian for steeper PSD indices.

Panel (b) on this figure shows example time series, PSDs and PDFs for integer PSD indices from $\Gamma=0$ to $\Gamma=3$. The time series corresponding to a PSD index of $\Gamma=0$ is effectively white noise, and the PDF functional form is approximately Gaussian. For $\Gamma=1$, there is less rapid variability in the time series although the PDF does not appear to deviate significantly from a normal distribution. In the more extreme example cases of $\Gamma=2$ and even more so for $\Gamma=3$, the time series has even less structure and the PDF has almost a double peaked structure. In the case of $\Gamma=3$, the time series is almost completely dominated by the large amount of power concentrated at the lowest frequencies. The long wavelengths corresponding to these frequencies largely determine the light curve properties. The corresponding example PDF for $\Gamma=3$ is clearly significantly different from having a Gaussian or lognormal functional form. These properties were common for time series generated from power-law PSDs with these spectral indices.

This is further explored in Fig. \ref{fig:nonStat}, which demonstrates that short light curves obtained from the same simulated long lightcurve and therefore have been created from the same PSD may not necessarily be determined to have the same PDF. In this figure, the user-defined PSD for the long lightcurve takes a power-law form with a red noise spectral index of $\Gamma=2$. The figure illustrates that the three PDFs corresponding to the three short time series drawn from the long lightcurve have different functional forms; the green dashed light curve has a PDF with a nearly flat distribution, the red dotted light curve exhibits a PDF with a single large peak and the blue dot-dashed time series PDF looks bi-modal. This figure highlights the difficultly in correctly measuring a true PDF for a source of which limited time series data is available. Fig. \ref{fig:nonStat} illustrates than the divergence from normally distributed fluxes shown in Fig. \ref{fig:PSDev} can be attributed to weak non-stationarity, where the variance of individual light curve segments is not constant in time but instead varies about a mean value determined by the underlying PSD \citep{UMV:2005}. 

\section{Functional Form of mean $p$-value dependence on PSD index}\label{sec:func}

To quantify the impact that the PSD spectral index has on TK95 time series simulations, we performed an empirical fit to the data shown in Fig. \ref{fig:PSDev}a. This functional form of this curve is given by that of a modified sigmoid, namely,

\begin{equation}
    f(\Gamma) = \frac{\alpha \exp\left(\beta - \eta \Gamma \right)}{1 + \exp \left(\beta - \eta \Gamma \right)},
    \label{eq:func}
\end{equation}
where $\Gamma$ represents the PSD spectral index assuming a PSD defined by Eqn. \ref{eq:PSD} and the parameters $\alpha$, $\beta$ and $\eta$ are free. The parameter $\alpha$ effectively normalises the curve, $\beta$ represents an x-axis translation and $\eta$ quantifies the steepness of the decline in average $p$-value, which becomes more rapid as $\eta$ increases. 

In the following section, we describe further simulations undertaken to understand the influence of the frequency range of the PSD on the $p$-value vs $\Gamma$ sigmoid curve and discuss their impact on the best fit of the functional form described by Eqn. \ref{eq:func}.

\section{Further Tests}\label{sec:MoreTests}
\begin{figure}{\includegraphics[width=\columnwidth]{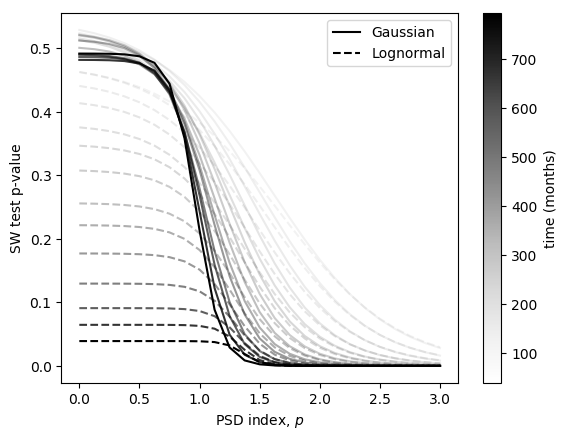}} 
\caption{Figure showing the evolution of the best fit described by Eqn. \ref{eq:func} as in Fig. \ref{fig:PSDev} with respect to the length of the time series. Artificial light curves were generated via the TK95 method and all have intrinsically Gaussian PDFs. They were tested for normality in linear space (solid lines) and log space (dashed lines) via SW tests. It can be seen that, for $\Gamma \leq 1$, the normalisation is roughly constant for the former, but the artificial time series PDFs are on average less likely to be lognormal for longer light curves, indicated by the declining SW test $p$-value associated with the test for normality in log space. Additionally, the deviations from normality occur at a more rapid rate for longer time series. This figure illustrates the effect changing $\nu_{\rm{min}}$ has on TK95 time series produced from a power law PSD. }
\label{fig:FitvsDataLength}
\end{figure}

\begin{figure}
    \centering
    \includegraphics[width=\columnwidth]{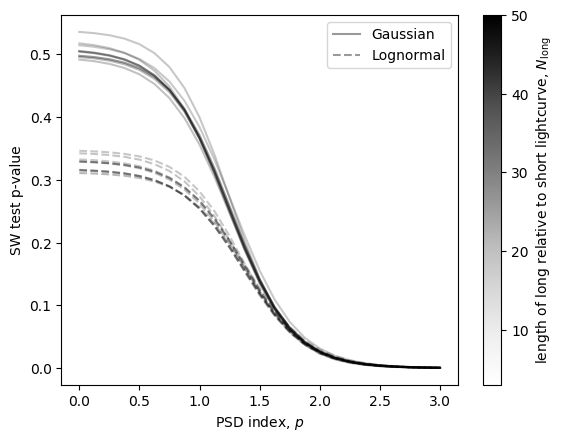}
    \caption{The effect of red-noise leakage as a function of PSD index. Here, artificial time series each containing 128 elements and with fixed $\nu_{\rm{max}}$ were generated from power-law PSDs via the TK95 method. The length of the long lightcurve, $N_{\rm{long}}$, from which a shorter lightcurve of the desired length was extracted, was varied to investigate the effect of red-noise leakage, with the corresponding noisy power-law PSDs extending to lower frequencies for larger $N_{\rm{long}}$. As leakage more strongly reduces the power contained in low frequency components and transferring it to high frequency components, this effect is more prominent in low $\Gamma$ regions where high frequency components contribute significant power to the time-series. }
    \label{fig:FitsvNlong}
\end{figure}

\begin{figure}{\includegraphics[width=\columnwidth]{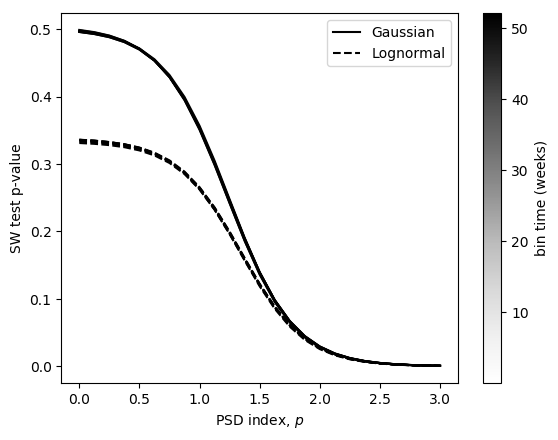}} 
\caption{As Fig. \ref{fig:FitvsDataLength} but changing the timescale associated with the smallest bin. Unlike Fig. \ref{fig:FitvsDataLength}, it can be seen that the bin time has no effect on the overall dependency of the produced time series with respect to normality. This is because although $\nu_{\rm{max}}$ is changed, the relative power contributions from each frequency element in the PSD remain the same.} 
\label{fig:FitvsBinsz}
\end{figure}

Having analytically motivated and demonstrated the decline of normally distributed TK95 time series beyond $\Gamma \approx 1$, we decided to undertake further experiments to establish whether the characteristic shape outlined in Eqn. \ref{eq:func} is influenced by other properties of TK95 simulations to test each of them. Specifically, we wished to establish if the position and severity of the break is influenced by the following:

\begin{itemize}
    \item The total length of each artificial time series. As a longer time series samples lower frequencies, this effects $\nu_{\rm{min}}$.
    \item The amount of red noise leakage in the time-series.
    \item The binning of, or the smallest timescale associated with, the time series. This changes $\nu_{\rm{max}}$.
\end{itemize}

To look into this, further simulations were carried out. To address the first point, the time binning was kept constant as monthly but the total length of the time series was increased incrementally up to 800 elements, equivalent to $\sim 67$ years. The same simulation setup was taken, with a plot of mean $p$-value vs PSD index obtained for each time series length, with the mean taken from computing the SW $p$-value from 10,000 artificial TK95 light curves generated for each PSD index with intrinsically Gaussian PDFs. This plot was obtained for testing the mean $p$-values of the produced time series both in linear and log space, with a mean best fit parameter curve produced from these simulations by fitting Eqn. \ref{eq:func} and varying $\alpha$, $\beta$ and $\eta$.

The results of this test are shown in Fig. \ref{fig:FitvsDataLength}. It can be seen that although there is variation in the curve with respect to the total length of the time series, each curve still indicates that the generated artificial time series show, on average, increasing deviations from being normally distributed at higher PSD indices. However, whilst increasing the length of the time series has little effect on the shape of the curve for the linear space SW $p$-value for $\Gamma \leq 1$, the $p$-value testing normality in log space decreases as a function of time series length, corresponding to a reduction of the parameter $\alpha$. To some extent this is a trivial result as we would expect to recover the intrinsic PDF shape more easily when taking more samples as in the case of a longer time series, yet it highlights the need for caution when attempting to determine the intrinsic PDF for a source with a relatively small time series containing few data elements. This arises because the SW test has more power for larger sample sizes, thus returns a result more likely to correctly reject the null hypothesis in the lognormal case. 



Fig. \ref{fig:FitvsDataLength} was generated from simulations of long lightcurves 10 times the desired length with noisy power-law PSDs \citep{TK:95}. Accordingly, the PSD extends to lower frequencies as the length of the lightcurve increases. It follows that for a PSD index steeper than white noise, the ratio of power contained in the constant maximum sampled frequency (shortest sampled timescale) relative to the variable minimum frequency decreases as the minimum frequency becomes smaller. This is shown in Fig. \ref{fig:FitvsDataLength} by the steeper transition at $\Gamma \approx 1$ (corresponding to an increase in the best fit parameter $\eta$ in Eqn. \ref{eq:func}) for longer time series as there is relatively less power contained at high frequencies for longer time-series relative to shorter ones as the former have power-law PSDs which extend to a lower minimum frequency. This allows the higher frequency elements to have a greater influence on shorter time series, reflected in the less rapid decline of the mean \pval in Fig. \ref{fig:FitvsDataLength} relative to those for longer time series.

The effects of red-noise leakage can be mitigated by simulating time-series from PSDs that extend to a lower frequency range and drawing a lightcurve of the desired length from a longer time-series \citep{2003MNRAS.345.1271V}. Therefore, to quantify the effect of this another test was undertaken, this time varying the length of the long lightcurve, $N_{\rm{long}}$, from which the desired short time-series (of 128 elements) was extracted. For higher values of $N_{\rm{long}}$, the (power-law) PSD extended to a proportionally lower frequency range. Results are shown in Fig. \ref{fig:FitsvNlong}. It is immediately obvious that at steep PSD indices ($\Gamma \gtrapprox 1.5$) that all of the curves show great consistency. This is because at steep PSD indices, the lower frequency components dominate the time-series irrespective of leakage. Leakage more strongly reduces the power contained in lower frequency components in the time-series, indicated by the divergence of the curves for $\Gamma \lessapprox 1.5$. Generally, the curves have lower mean \pvals for larger $N_{\rm{long}}$ when the effects of leakage are smaller. This is because red-noise leakage more strongly reduces the power contained in low frequency components, effectively making the PSD power-law index less steep and changing the \pval to be consistent with that expected from a shallower PSD index.

A third test was undertaken, this time varying the minimum sampling time, which changes the maximum frequency in the PDF. The range explored was from 1 hour to 1 year, with common sampling such as daily and monthly time bins investigated. The results of this are shown in Fig. \ref{fig:FitvsBinsz}. It is clear that changing this parameter has almost no effect on the parameterisation curves at all, as they are all close to being exactly superimposed. Changing the highest PDF frequency therefore has a negligible effect on the properties of TK95 time series simulations. This is because the relative power in each frequency component is the same in this test, i.e. the time series has the same number of elements regardless of the sampling time. This effectively translates the PSD on the y-axis, whereas in the previous test increasing the length of the time series extended the frequency range and light curve length thus changing the relative distribution of power into frequencies sampled by each artificial time series.

Throughout our tests, if the sample size of the time series is $\geq 30$, there is great consistency in the value of the mean $p$-value for $\Gamma \lessapprox 1$ when the light curve PDFs are consistent with a Gaussian functional form. This is shown in Figs. \ref{fig:PSDev}a, \ref{fig:FitvsDataLength} and \ref{fig:FitvsBinsz} and is equivalent to stating that $\alpha$ in eqn. \ref{eq:func} is approximately 0.5. This result is expected as the distribution of $p$-values should be flat if the null hypothesis cannot be rejected \citep{rice2006mathematical,Murdoch:2008}, i.e. TK95 time series PDFs are consistent with having normally distributed fluxes for $\Gamma \leq 1$. This result is perhaps more intuitively understood when considering the mean $p$-value when testing for normality in log space, as depicted in Fig. \ref{fig:FitvsDataLength}. Here, the $p$-value decreases as the length of the time series increases, implying that the $p$-value distribution becomes skewed to favour lower numbers as a greater number of TK95 simulations reject the null hypothesis. In this figure, as the time series length increases so does the statistical power, and  a greater proportional of time series realisations correctly have the null hypothesis, i.e. that they are normally distributed in log-space, rejected. Therefore the distribution of $p$-values becomes skewed as it favours values closer to zero, which is reflected in the lower mean values.

\section{False Positive Rate for PDFs}\label{sec:falsePos}

\begin{figure}{\includegraphics[width=\columnwidth]{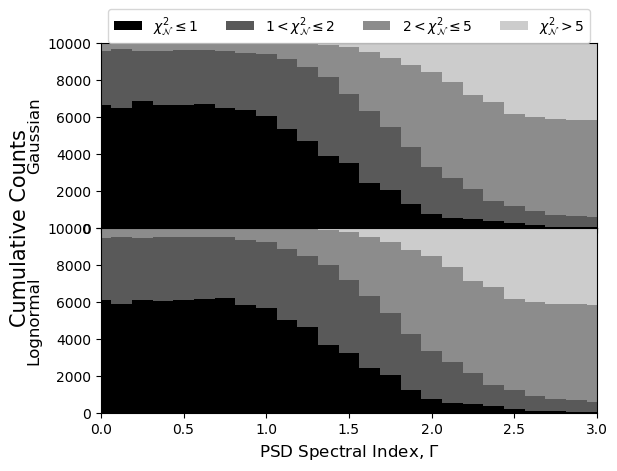}}
\caption{Based on 10,000 TK95 generated artificial times series each with 128 elements (each being one month) and intrinsic Gaussian PDFs. It can be seen that for $\Gamma \leq 1$, a Gaussian is, on average, an appropriate functional form for describing the PDF when evaluated with a $\chi_{\mathcal N}^2$ test. In this $\Gamma$ range, a lognormal also often provides a good fit to the PDF, though the quality of the fit is on average poorer than the Gaussian. It can be seen that the quality of both fits declines rapidly beyond $\Gamma=1$, after which neither distribution is likely to accurate describe the PDF. This is as a result of non-stationarity of the PSD as all segments of the long simulated time series have different properties \citep[see also][]{2019MNRAS.485..260A}. }
\label{fig:chibothTK}
\end{figure}

\begin{figure}{\includegraphics[width=\columnwidth]{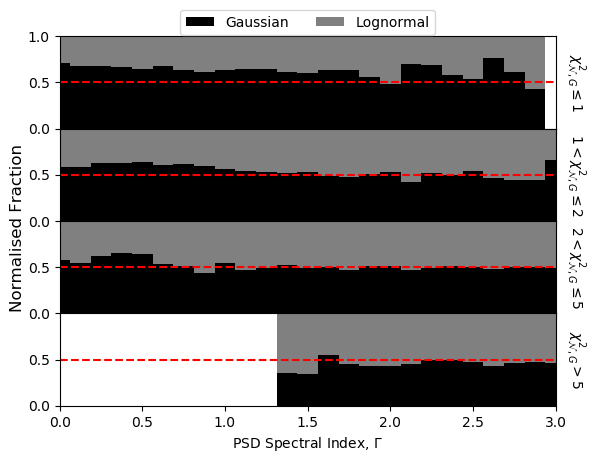}}
\caption{Comparison of Gaussian vs lognormal fits to the PDFs of artificial time series. A logrnormal model was deemed a better fit for $\chi_{\mathcal{N}, \rm{LN}}^2 < \chi_{\mathcal{N}, \rm{G}}^2$, with the fraction of PDFs preferred indicated by the grey portion of the bars. The subplots divide the fits based on the values of $\chi_{\mathcal{N}, \rm{G}}^2$. Based on 10,000 TK95 generated artificial times series each with 128 elements (each being one month) and intrinsic Gaussian PDFs. For clarity, only indices for which at least 30 convergent fits were obtained are displayed with a bar in the figure. It can be seen in these simulations a normal distribution typically better describes the shape of the PDF than a lognormal function for low $\chi_{\mathcal N}^2$ and when $\Gamma<1$. However, above this index, especially when both fits are poor, the false positive fraction is around 50\%, which is indicated by the dashed horizontal line.}
\label{fig:falseP_norm}
\end{figure}

\begin{figure}{\includegraphics[width=\columnwidth]{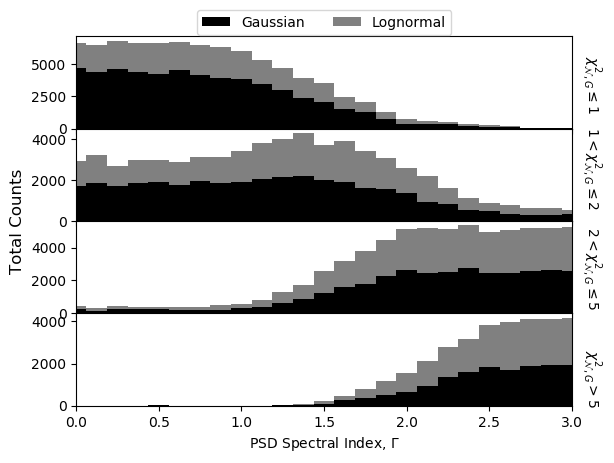}}
\caption{Based on 10,000 TK95 simulated times series each with 128 elements (each being one month). This figure shows the raw counts in each \chisq band, from which the normalised fraction in Fig. \ref{fig:falseP_norm} were calculated. It can be seen that good fits to the PDF from a Gaussian or lognormal function are more likely to be obtained when the PSD index is low. }
\label{fig:falseP_unnorm}
\end{figure}

It should also be stated that there are numerous ways of estimating the PDF from data, including non-parametric ones \citep[see][for a summary]{PDF:nonparametric}. Each has its own distinct advantages and disadvantages, but a comprehensive discussion is beyond the scope of this paper. In astronomy literature, the functional form of the PDF is often evaluated by binning the time series and fitting to the resultant histogram \citep[e.g.][]{2016MNRAS.460.1790G,2017ApJ...849..138K,2018RAA....18..141S}, with the comparison of different functional forms often completed using a \chisq test. So far, we have assessed deviations from normality in the PDFs of simulated TK95 series using the SW test as our metric. The most popular, physically motivated functional forms of the PDF found in the literature are Gaussian and lognormal, so we first extend the previous work to establish whether Gaussian TK95 PDF light curves can be erroneously measured as lognormal. 

Fig. \ref{fig:chibothTK} shows the distributions of \chisq values for an ensemble of 10,000 TK95 generated time series. This reflects the result Fig. \ref{fig:PSDev}, although this figure shows that the best fit of a lognormal function to the histograms gets progressively worse on average for $\Gamma \gtrapprox 1$, much like for the Gaussian fit. This is an important result as it captures the essence of statistical behaviour arising from loss of strict stationarity. Although these simulations deviate from being normally distributed at steep PSD indices, they do not show any preference for the lognormal models either. This is because for steep PSDs, the divergence of power at long timescales, implies divergence of the corresponding variance. And this renders higher moments of the PDF ill-constrained (and varying significantly from realisation to realisation). As a by-product, this diminishes the chance of falsely obtaining an "incorrect result", which corresponds to identifying distribution as lognormal. This is owing to the binary choice, "normal vs lognormal" provided. The true PDF of course, could be more complex and also different from both choices. However extracting this general form which is well-motivated physically, from the data is challenging when multiple, complex processes are at play.

Fig. \ref{fig:falseP_norm} compares the \chisq values for Gaussian fits to those for lognormal fits for the same PSD index, for a range of values. The four bands shown are somewhat arbitrary, with the \chisq range corresponding to the Gaussian fits only. A lognormal model was deemed a better fit if $\chi_{\mathcal{N}, \rm{LN}}^2 < \chi_{\mathcal{N}, \rm{G}}^2$. These cases are indicated by the grey portion of each bars, whereas $\chi_{\mathcal{N}, \rm{G}}^2 < \chi_{\mathcal{N}, \rm{LN}}^2$ are shown in black. Here, the proportion of each bar that is grey is the false positive fraction, i.e. the percentage of time in which a lognormal PDF fit is erroneously preferred over a Gaussian functional form. It can be seen for this example that the incorrect intrinsic PDF is the preferred fit in $\gtrapprox 60\%$ of cases, with a false positive unsurprisingly less likely to be determined in the event of a good \chisq fit. Fig. \ref{fig:falseP_unnorm} shows the number of counts used to normalise each bar in Fig. \ref{fig:falseP_norm}, and also shows the evolution of the \chisq distribution. It can be seen that the quality of fit of a Gaussian function to the histogrammed PDF worsens significantly as the PSD index steepens, verifying our result from Section \ref{sec:method}. This result is important as it demonstrates the use simulated time series can have in testing whether an observed time series has a correctly measured PDF. 

We have investigated the suitability of TK95 simulated time-series for reproducing normally distributed artificial time-series. Before applying these results to real time-series, we briefly outline some additional caveats of generating artificial time-series.


One such caveat of generating artificial time-series is that they are unable to reproduce the same skewness as observed for real time-series. This has been showed by calculating the bicoherence, which quantifies the coupling between variations at different timescales (and in principle can be used to distinguish between linear, Gaussian processes and non-linear processes, which can produce non-Gaussian PDFs), for real and artificial time series \citep{2002MNRAS.336..817M,UMV:2005}. Presently, no such method for the generation of artificial time-series exists which can include this feature, thus it is not accounted for here.

An additional caveat is that real time series can show strong non-stationarity, which manifest themselves as an additional distortion of the flux PDF. This has been demonstrated by monitoring the X-ray variability of the narrow-line Seyfert 1 (NLS1)
galaxy IRAS 13224-3809 \citep{2019MNRAS.482.2088A,2019MNRAS.485..260A}, where the PSD is strongly non-stationary and the PDF deviates from the expected lognormal functional form. As the computation of the PSD becomes more challenging for smaller data sets, this effect is unlikely to be a major issue for the majority of time series, and can be tested for using the methods described in \citet{2003MNRAS.345.1271V} and \citet{2019MNRAS.482.2088A}. 

In the remainder of this paper we apply our results to real blazar data. The objective is to prescribe a false positive fraction, specifically what fraction of the time one can expect to measure a PDF that is not intrinsically the PDF of the object of interest. When assessing real data, it is difficult to ascertain the true PDF functional form, therefore simulated time series with known PDFs and PSDs are a powerful tool which can be utilised to derive a false positive fraction. In this section, we assume that sources can have PDFs of a Gaussian or lognormal functional form, and ignore other possibilities. To begin with, let us assume that a PDF consistent with a Gaussian functional form is observed. We wish to know what is the probability that the true source PDF is Gaussian given we have measured it to be so, \pggpdf. Bayes' theorem \citep[e.g.][]{sivia2006data} tells us this is,

\begin{equation}
  P({\rm G}|{\rm G}_{\rm{PDF}}) = \frac{P({\rm G}_{\rm{PDF}}| {\rm G}) P({\rm G})}{P({\rm G}_{\rm{PDF}})},
  \label{eq:PG}
\end{equation}
where \pg is the prior probability that the true PDF is Gaussian. We wish to evaluate the probability of obtaining a Gaussian PDF, \pgpdf, which under our assumption that the true PDF must be either Gaussian or lognormal can only be obtained in two ways. These are either from correctly measuring a Gaussian PDF from a source with a true Gaussian PDF, or incorrectly measuring a Gaussian PDF from a source with an intrinsic lognormal PDF. Using $+$ or $-$ to denote either a correct or incorrect PDF measurement, and \pln as the probability for a true lognormal PDF allows us to write \pgpdf as,

\begin{equation}
  P({\rm G}_{\rm{PDF}}) = P(+|{\rm G})P({\rm G}) + P(-|{\rm LN})P({\rm LN}),
  \label{eq:pgpdf}
\end{equation}
which, when substituted into Eqn. \ref{eq:PG} gives,
\begin{equation}
  P({\rm G}|{\rm G}_{\rm{PDF}}) = \frac{{\rm P}(+|{\rm G}) P({\rm G})}{P(+|{\rm G})P({\rm G}) + P(-|{\rm LN})P({\rm LN})},
  \label{eq:GGpdf}
\end{equation}
which defines the true positive fraction. Similarly, we can calculate the false positive fraction as the probability of measuring a lognormal PDF given the true PDF is Gaussian as,
\begin{equation}
    P({\rm LN}|{\rm G}_{\rm{PDF}}) = \frac{P(-|{\rm LN}) P({\rm LN})}{P(+|{\rm G})P({\rm G}) + P(-|{\rm LN})P({\rm LN})}.
    \label{eq:LNGpdf}
\end{equation}
Eqns. \ref{eq:GGpdf} and \ref{eq:LNGpdf} apply to sources with intrinsically Gaussian PDFs. In a similar fashion, we can calculate the true and false positive fractions for intrinsically lognormal sources using, 
\begin{equation}
  P({\rm LN}|{\rm LN}_{\rm{PDF}}) = \frac{P(+|{\rm LN}) P({\rm LN})}{P(+|{\rm LN})P({\rm LN}) + P(-|{\rm G})P({\rm G})},
  \label{eq:LNLNpdf}
\end{equation}
\begin{equation}
  P({\rm G}|{\rm LN}_{\rm{PDF}}) = \frac{P(-| {\rm G}) P({\rm G})}{P(+|{\rm LN})P({\rm LN}) + P(-|{\rm G})P({\rm G})}.
  \label{eq:GLNpdf}
\end{equation}
In these formulae, \pg and \pln are the priors, while the variables \posg, \negg, \posln and \negln can be used from the time series simulation techniques outlined in the previous sections. We demonstrate this in the remainder of the paper, demonstrating the use of this technique with a worked example.

\subsection{A Worked Example}

\subsubsection{Fermi Analysis}

For our example object, we used a bright source because if the source in question is often only weakly or un-detected, it may skew the shape of the PDF as flux values by definition cannot be negative. In this example we investigate the blazar PKS2155-304, for which we produce a time series in the energy range 100 MeV - 300 GeV by analysing publicly available {\sl Fermi} LAT data.

\begin{figure}{\includegraphics[width=\columnwidth]{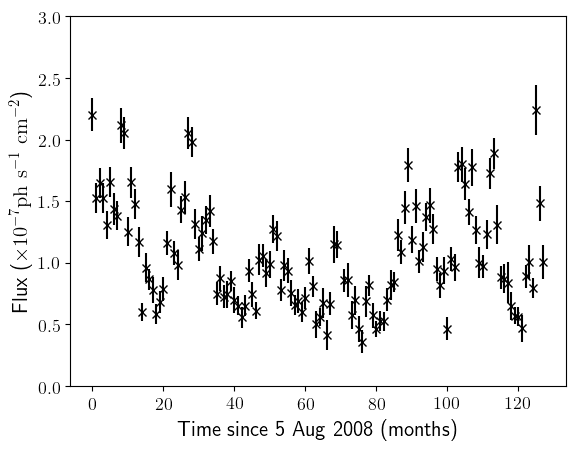}}
\caption{Monthly time binned {\sl Fermi} LAT light curve for the blazar PKS2155-304. The energy range is from 100 MeV to 300 GeV. The light curve shows the 128 30-day intervals between Aug 5 2008 and Feb 9 2019.}
\label{fig:2155LC}
\end{figure}

To produce a time series, we first downloaded a photon file centred on the target source and containing all detected photons within a 15$^{\circ}$ radius that had been emitting during the first $\approx 10$ years of {\sl Fermi} LAT observations, beginning 5 August 2008 and ending 9 Feb 2019. For the same interval, we downloaded the spacecraft file which details the relative orientation of the LAT to each source and is necessary for the analysis. The detected photons were then divided into 128 30-day time bins to ensure the source was significantly detected ($TS \geq 25$) in each light curve interval. The data corresponding to each time bin were subject to an unbinned analysis using the P8R2\_SOURCE\_V6 instrument response functions, accounting for the Galactic and isotropic background photons by incorporating the models gll\_iem\_v06.fits and iso\_P8R2\_SOURCE\_V6\_v06.fits, of which the normalisation of the latter was left as a free parameter. These models can be downloaded from the {\sl Fermi} LAT data server\footnote{\url{https://fermi.gsfc.nasa.gov/ssc/data/access/}}. The input model map for our analysis froze all parameters for Fermi 3FGL catalogue sources \citep{2015LATcat} $>10^{\circ}$ from our target of interest as the point-spread function (PSF) of the Fermi LAT at 100 MeV is $\approx 3.5^{\circ}$ and decreases at higher energies \citep{aharonian2013astrophysics}, so we do not expect photons from these sources to contaminate data from PKS2155-304. Sources within $10^{\circ}$ of PKS2155-304 were modelled with their normalisations as free parameters, freezing other parameters to their 3FGL catalogue values \citep{2015LATcat} as we are not interested in spectral information and only wished to produce a time series. The best fit was found using the NEWMINUIT algorithm \citep{James:1994vla}, which returned flux values and uncertainties based on predicted photon counts associated with each source. The produced time series can be seen in Fig. \ref{fig:2155LC}.

\subsubsection{Determining the False Positive Fraction}

\begin{figure}{\includegraphics[width=\columnwidth]{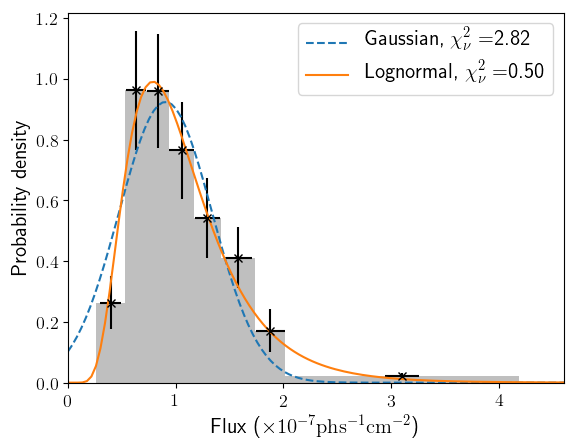}}
\caption{The PDF for PKS2155-304 corresponding to the time series shown in Fig. \ref{fig:2155LC}. The vertical error bars were calculated assuming the error is proportional to the number of counts in each bin. The histogram was generated using the condition that the mean error of data in each bin, indicated by the horizontal error bars, could not exceed the bin width. Additionally, each bin was required to have $\geq 5$ data points. It can be seen that the histogram functional form is better described by a lognormal distribution as opposed to a Gaussian function.}
\label{fig:2155PDF}
\end{figure}

\begin{figure}{\includegraphics[width=\columnwidth]{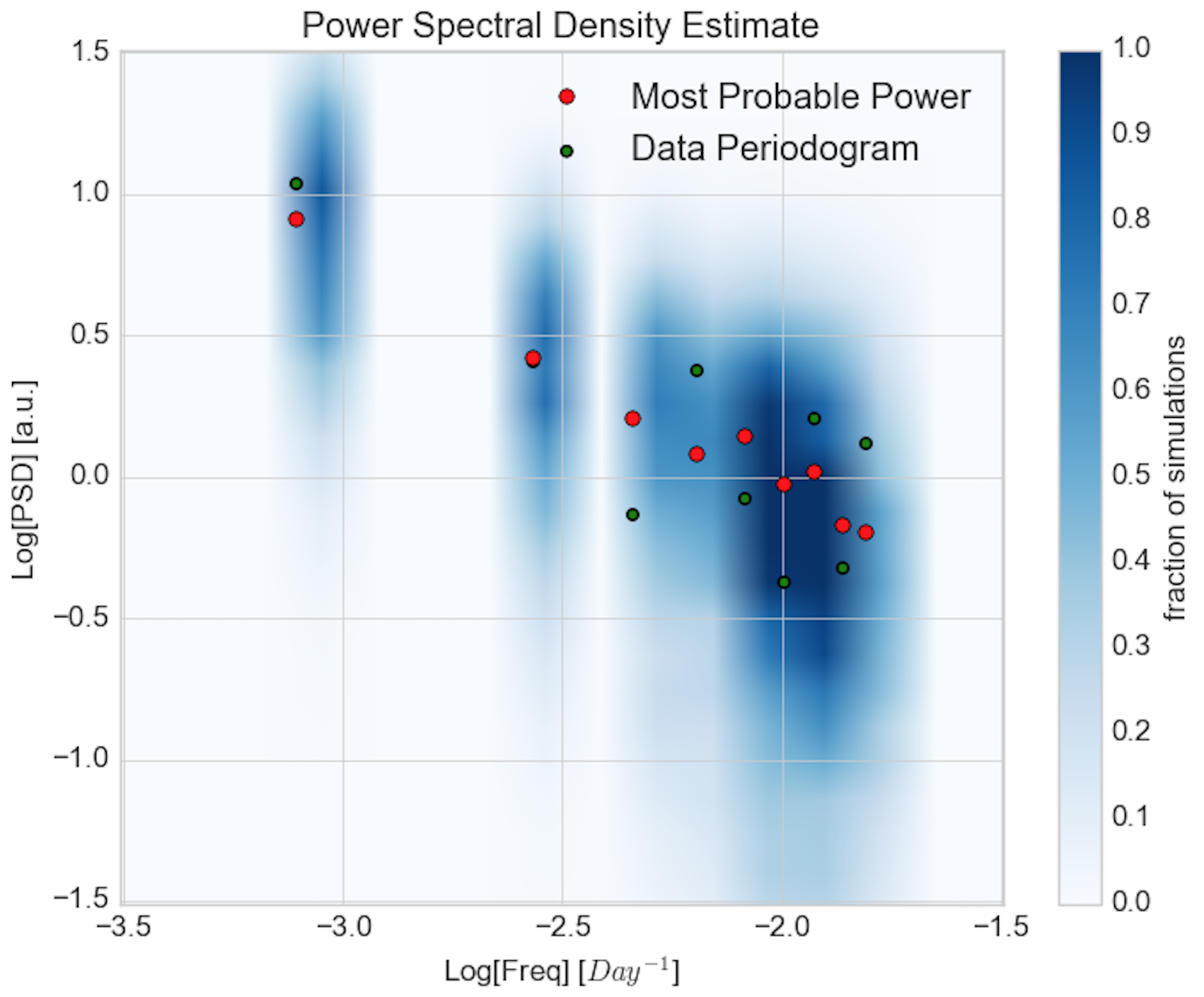}}
\caption{The PSD estimate of Fermi lightcurve of PKS 2155-304 using modified Timmer and Koenig simulations with exponential transform for lognormality is shown. The colormap represents the simulated power in each frequency bin for PSD indices from 0.2 to 2.8 in steps of 0.2. The color scale represents the fraction of simulations within a power bin for a frequency range. The red represents the ``most probable power" in a given frequency bin for every index being equiprobable. The green points represent the periodogram of the observed lightcurve. The PSD is consistent with pink noise.}
\label{fig:2155PSD}
\end{figure}

From the produced time series, both the PDF and PSD can be estimated. To create a PDF, we bin the light curve in accordance to pre-existing practices in the literature, whereby we account for the size of the error bars on the data points \citep[e.g.][]{2017ApJ...849..138K} and ensure that each bin in the PDF is wider than the mean error bar of data points within that bin. Additionally, we impose the condition that each bin must contain $\geq 5$ data points \citep[see][]{HandH} to prevent a small number of outliers from skewing the distribution which may erroneously lead to one PDF functional form being erroneously preferred. The resultant PDF is illustrated in Fig. \ref{fig:2155PDF}, and it can be seen that its functional form is better described by a Gaussian than for a lognormal. 


For our false positive calculation we choose to input the measured PSD spectral index of the time series in question into our TK95 simulations. Measuring the PSD index for real data can be achieved by taking a discrete Fourier transform of the time series such that $\mathcal{P} \propto |\mathcal{F}(t)|^2/N$  \citep[e.g.][]{TK:95}. From this the PSD spectral index can be estimated in different ways. Or equivalently there are different estimators. Two estimates are made here. First is to compare the chi-square, $\chi^{2}_{\rm{dist}}$ of the mean simulated power to that of the observed power $\chi^{2}_{\rm{obs}}$, which are given by $\chi^{2}_{\rm{dist},i} = \Sigma^{\nu=\nu_{\rm{max}}}_{\nu=\nu_{\rm{min}}} \frac{[P(\nu)_{\rm{sim},i} - \langle P(\nu)_{\rm{sim}} \rangle]^2}{[\langle \Delta P(\nu)_{\rm{sim}} \rangle]^2}$ for the ith simulation and $\chi^{2}_{\rm{obs}} = \Sigma^{\nu=\nu_{\rm{max}}}_{\nu=\nu_{\rm{min}}} \frac{[P(\nu)_{\rm{obs}} - \langle P(\nu)_{\rm{sim}} \rangle]^2}{[\langle \Delta P(\nu)_{\rm{sim}} \rangle]^2}$ respectively. The fraction of simulations where $\chi^{2}_{\rm{obs}} < \chi^{2}_{\rm{dis}t,i}$ is the fraction whose PSD estimate values are within the statistical spread of simulations. For PKS 2155-304, we find that for a range of indices from 0.2 to 2.8 with increments of 0.2, this fraction peaks at 1.2. In an alternate way, we can determine both the central value and uncertainty of the index. Here we calculate the mean power in each frequency bin over range of indices 
as shown in Fig. \ref{fig:2155PSD}. Note that since the cadence of the observations as well as the mean and variance of the lightcurve are imposed on the simulations, this mean power map or "mean PSD map" is not unbiased but carries information on the moments of the observed lightcurve.  From this mean "PSD map", we can compute the best fit value which for PKS2155-304 it is $\Gamma \sim 0.9 \pm 0.5$. This best fit is derived from the "most probable powers" in each frequency bin 
which are shown in red in Fig. \ref{fig:2155PSD}. The green points show the periodogram directly computed from the lightcurve. Fig. \ref{fig:2155PSD} shows that the observed lightcurve is roughly consistent being a realisation of a pink noise process. 

Previously we have shown that TK95 simulations are appropriate for constraining \posg, \negg, \posln and \negln in Eqns \ref{eq:GGpdf}-\ref{eq:GLNpdf}.
\posg, \negg, \posln and \negln can be determined in the same way as used to produce Fig. \ref{fig:falseP_norm}. Using the best fits to both the Gaussian and lognormal distributions fitted to the PDF, 10,000 artificial time series are generated for each. Using the results from our calculation of the PSD index, for each artificial time series we draw a value of $\Gamma$ from a normal distribution characterised with mean $\Gamma = 0.9$ and $\sigma=0.5$. Fig. \ref{fig:chibothTK} shows that in this regime, time series which are on average normally distributed can be quickly produced via the TK95 method. Similarly, lognormal time series can be obtained by exponentiating a normally distributed time series. One caveat here is that exponentiating the time series can change the time series PSD. \citet{UMV:2005} have investigated this effect (see Appendix B) and find that although slightly more power is present in high-frequency components, the overall effect is small for PSDs lacking sharp features, such as the power-law PSDs used in this example. Alternatively, the EMP13 algorithm may be used; it is more general with user defined PDF and PSD which are refined consistently with the data. However, it is naturally more complex and computationally expensive. In this case, we use TK95 as we in are in the $\Gamma\gtrapprox1$ regime so the flux distribution for each artificially generated time series will on average show a high degree of consistency with the true PDF. Each simulated time series has its PDF evaluated in exactly the same way as the real time series, using the same data binning as we are assuming error bars on each simulated time series data points to be correlated with that of the real data (although we do not explicitly calculate them). This allows  \posg, \negg, \posln and \negln to be determined.


\begin{figure}
\includegraphics[width=\columnwidth]{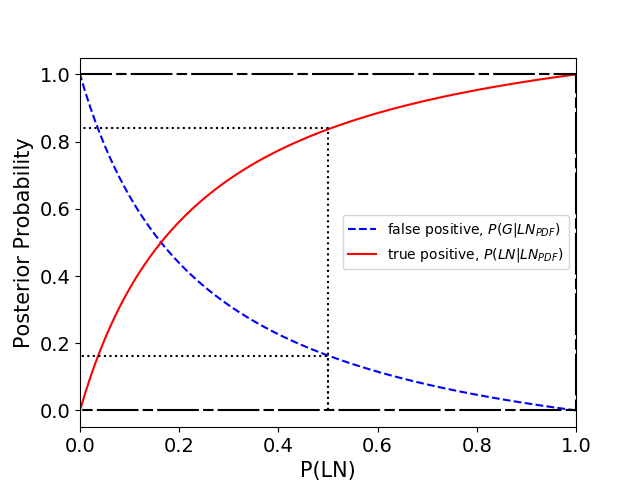}
\caption{Figure showing the variation of the posterior distributions for the PDF of $\gamma$-ray time series of PKS2155-304 being lognormally distributed as a function of the prior probability that the source has a lognormal PDF, $P({\rm LN})$. It was assumed that lognormal and normally distributed PDFs were the only options, such that the probability of a Gaussian PDF is $P({\rm G}) = 1 - P({\rm LN})$. It can be seen for our example of PKS2155-304, an intrinsically Gaussian PDF is only preferred if there is a high ($P({\rm G}) \gtrapprox 0.82$) prior probability that the PDF is Gaussian. The PDF and PSD for PKS2155-304 are shown in Figs. \ref{fig:2155PDF} and \ref{fig:2155PSD}. The dotted and double-dashed lines refer to choices of prior which are discussed in the text.}
\label{fig:probVprior}
\end{figure}


It remains to determine the prior probability that the PDF we are measuring is Gaussian (\pg) or lognormal (\pln). From our assumption that the PDF can only take one of these functional forms, $P({\rm LN}) = 1 - P({\rm G})$. Fig. \ref{fig:probVprior} shows the posterior distribution as a function of \pln, demonstrating the effect that varying \pln has on the posterior distributions for a true positive, indicated by the red solid line and defined by Eqn. \ref{eq:LNLNpdf}; and for a false positive, which is shown by the blue dashed line and defined in Eqn. \ref{eq:GLNpdf}. In this figure, we highlight two possible choices of prior.

The first of these utilises the principal of indifference, in which we assume the simplest possible non-informative (or flat) prior where \pg~$=$ \pln~$=0.5$. This is marked on Fig. \ref{fig:probVprior} using the black dotted line, and corresponds to values of \pglnpdf~$= 0.163$ and \plnlnpdf~$= 0.837$, indicating a preference for lognormality somewhere between 1 and 2 $\sigma$. 

Secondly, we compute a second example using priors weighted on the probability that our histogram is representative of the true PSD. To do this, we use the Bayesian inference to calculate the probability of obtaining each datum, $D_{\rm{i}}$, assuming the model with parameters $\theta_{\rm{i}}$, is correct. For the entire histogram PDF shown in Fig. \ref{fig:2155PDF}, this can be expressed as,

\begin{equation}
    P(D|\theta_{\rm{i}}) \propto \prod^N_i \exp{\left[ -\frac{ \left( D_{\rm{i}} - m_i(\theta_i)  \right)}{2\sigma_i^2} \right]},
    \label{eq:prior}
\end{equation}
where $m_i(\theta_i)$ are the model values and $\sigma_i$ are the size of the uncertainties. It is assumed that measurements of each data point are normally distributed about the mean model value, and $N$ is the number of bins in the PDF. The total likelihood is the product of the likelihoods of each bin value being correct. We apply Eqn. \ref{eq:prior} to the normal and lognormal models for the PDF of PKS2155-304 and re-normalise under our assumption that the true PDF must be either Gaussian or lognormal. For PKS2155-304, we obtain $P({\rm G}) = 0.000587$ and $P({\rm LN}) = 0.999413$, where the low probability terms in \pg come from the bins at either end of the histogram, which are clearly a poor fit to the Gaussian model in Fig. \ref{fig:2155PDF}. The result of these much more extreme values on our final results are that the false positive fraction becomes \pglnpdf~$= 0.000115$ and the true positive \plnlnpdf~$= 0.999885$, with the latter of these $> 3 \sigma$. It is clear that this prior leads to a much stronger preference for lognormality, but we stress this is largely because for this prior the probability that the histogram bins on the tail of the PDF are very small assuming a Gaussian model, and we have not tested other functional forms.

Fig. \ref{fig:probVprior} is a key and informative result as it demonstrates the difficulty associated with being able to conclusively say that an individual source has a particular intrinsic PDF distribution, irrespective of the prior chosen. In our example of PKS2155-304, we have shown that a simple histogram PDF exhibits a preference for a lognormal functional form, yet we cannot say we are confident that is a correct measurement of the true intrinsic PDF to much more than $3\sigma$. Even so, Fig. \ref{fig:probVprior} shows an overall preference of PKS2155-304 to have a lognormally distributed PDF as we are only likely to obtain a Gaussian PDF measurement if our lognormal prior has a value $P(LN) \lessapprox 0.18$. The point where \pglnpdf~$=$~\plnlnpdf~ is therefore an important diagnostic for determining a false positive rate for an individual time series. We recommend undertaking artificial light curve simulations using the method outlined above to determine how likely obtaining the inferred result is.


In summary our false positive prescription may be outlined as follows:

\begin{itemize}
  \item Use the observed time series to compute a PSD and PDF, fitting the desired functional forms to each and evaluating them with a statistical test such as a reduced $\chi^2$. 
  
  \item Determine the priors $P({\rm G})$ and $P({\rm LN})$. In our example we have computed results for a flat prior and a histogram-weighted prior.

  \item Produce $N$ artificial time series from each distribution (Gaussian and lognormal) used to best fit the data PDF, including the measured PSD spectral index. We recommend using TK95 simulations to produce normally and lognormally distributed time series (normally distributed time series can be exponentiated to create lognormal time series), although this approach is only able to reliably produce time series with the correct PDF functional form when the PSD index is $\Gamma \lessapprox 1$. 
  \item As the true PDF functional form is known for the artificial time series, this will give $P(+|{\rm G}), P(-|{\rm G}), P(+|{\rm LN})$ and $P(-|{\rm LN})$, where $+$ and $-$ symbolise correct or incorrect measurement of inherently Gaussian (${\rm G}$) or lognormal (${\rm LN}$) PDFs.
  \item Depending on whether a Gaussian or lognormal PDF is measured from the time series, evaluate either equations \ref{eq:GGpdf} and \ref{eq:LNGpdf} or equations \ref{eq:LNLNpdf} and \ref{eq:GLNpdf} as appropriate to obtain positive and false positive fractions.
\end{itemize}




\section{Summary and Conclusions}

We have presented time series simulations based on the method of \citet{TK:95}, and investigated the relationship between the input PSD and properties of the output PDF. We initially began by assuming a simple power law PSD and investigated how this affected the average $p$-value returned from a Shapiro-Wilk statistical test, which tests for normality of the time series. We have used these tests to create a false positive prescription rate for evaluating the likelihood of the correct measurement of the PDF functional form for astrophysical sources. Our results may be summarised as follows:

\begin{itemize}
    \item PDFs of Timmer \& Koenig time series simulations are more likely to deviate from having normally distributed fluxes as the PSD index steepens. This is analytically expected because there is progressively less power in the higher frequency components relative to the lower frequencies. Beyond PSD index $\Gamma \approx 1$, artificial time series corresponding to a power law PSDs are increasingly dominated by contributions from these lower frequencies many of which have wavelengths longer than the desired observation time span. 
    \item The mean $p$-value from a Shapiro-Wilk test on TK95 simulated time series is roughly constant for PSD indices $\lessapprox 1$, but sharply begins to decrease above this index, eventually reaching $\sim 0$ implying artificial power law time series with $\Gamma >> 1$ on average reject the null hypothesis that the data is normally distributed.
    \item The characteristic fiducial fit to this curve is given by a sigmoid function, specifically, \\ \mbox{$f(\Gamma) = \alpha \exp\left(\beta - \eta \Gamma \right)/\left[1 + \exp \left(\beta - \eta \Gamma \right)\right]$}, where $\Gamma$ is the PSD spectral index and $\alpha$, $\beta$ and $\eta$ are free parameters. 
    \item In general it is difficult to distinguish between lognormal and Gaussian PDF functional forms for time series data, even more so for relatively small ($N\lessapprox 100$) data sets. We therefore recommend that the calculation of these for individual sources is accompanied by a false positive rate, the parameterisations of which are deduced from generating large numbers of artificial time series with known PDF distributions.
    \item As an example, we show that the {\sl Fermi} LAT $\gamma$-ray light curve of the blazar PKS2155-304 shows evidence of lognormality to 83.66\% when using a flat prior (assuming that an intrinsic Gaussian or lognormal PDF or equally likely). This increases to 99.99\% when weighting the priors using a Bayesian inference to evaluate the probability that the data points are correct assuming each model is.

\end{itemize}


\begin{center}
\textbf{Acknowledgements}
\end{center}

PJM acknowledges support from a Hintze Scholarship. NC acknowledges kind support by the AvH foundation and MPIK during development of this project. GC acknowledges support from STFC grants ST/N000919/1 and ST/M00757X/1 and from Exeter College, Oxford. This work was supported by the Oxford Hintze Centre for Astrophysical Surveys which is funded through generous support from the Hintze Family Charitable Foundation.




\bibliographystyle{mnras}
\bibliography{PSDbib}

\bsp	
\label{lastpage}
\end{document}